\documentclass[pra, twocolumn, showpacs]{revtex4-1}
\usepackage{float}
\usepackage{bbm}
\usepackage{epsfig}
\usepackage{epstopdf}
\usepackage{graphicx, subfigure}
\usepackage{amsmath,amssymb}
\usepackage{color}
\usepackage{hyperref}
\usepackage[capitalize]{cleveref}
\begin{document}

\title{Entanglement dynamics of two Ising-coupled qubits with nonperpendicular local driving fields}

\author{F.~A.~Calderon-Vargas$^{1,2}$}
\email{f.calderon@vt.edu}
\author{J.~P.~Kestner$^1$}
\affiliation{$^1$Department of Physics, University of Maryland Baltimore County, Baltimore, Maryland 21250, USA\\
$^2$ Department of Physics, Virginia Tech, Blacksburg, Virginia 24061, USA}

\begin{abstract}
We present an approximate analytical solution to the dynamic equation of two Ising-coupled qubits with oscillating classical control fields that are nonperpendicular to the static fields. This is a situation that has recently arisen in some solid-state experiments. With our solution we derive the analytical expressions for the local invariants as well as the local rotations needed to isolate a purely nonlocal gate. This determines the set of parameters that are required to generate any entangling gate. Moreover, we use our results to describe a recent experimental work on capacitively coupled singlet-triplet qubits in GaAs and discuss possible differences for a similar device in silicon.
\end{abstract}


\maketitle

\section{Introduction}
Examples of two-level quantum systems coupled to an oscillating field can be found in many areas of quantum physics, from quantum optics \cite{Jaynes1963,Shore1993,Raimond2001} to solid-state physics \cite{Hynek2002,Wallraff2004,Schuster2007,Hughes2008,Nichol2016}, and having an analytical expression for the evolution operator is relevant for obvious reasons. The rotating-wave approximation (RWA) is a widely used approach that provides an approximate analytical solution. The RWA is also used to analytically determine the quantum logic operations that are accessible by an oscillating Hamiltonian \cite{Zhang2005,Geller2010}. This approximation is mostly used in qubits whose oscillating control field is perpendicular to its static field, where the effect of the RWA is to produce a Hamiltonian that is time-independent in the rotating frame. However, not all systems present this perpendicularity between fields \cite{Goorden2003,Hausinger2010}. Reference~\cite{Zhang2005} discusses the proper use of the RWA to find an analytical solution to the dynamic equation of a single qubit with non-perpendicular fields.\\
\indent In recent experiments with capacitively coupled singlet-triplet qubits\cite{Nichol2016}, where the oscillating control fields are nonperpendicular to the static fields, the time-independent Hamiltonian is obtained using the RWA. However, this approximation is applied after neglecting the static field's component parallel to the oscillating field in order to recover perpendicularity between fields. While that simplified analysis gives quick insight, a more precise treatment that considers the full static field is necessary to fully capture the system's dynamics.\\
\indent In this work, we present an analytical solution to the dynamic equation of two Ising-coupled qubits with nonperpendicular fields followed by a characterization of the entangling properties of the two-qubit gates accessible by the system's Hamiltonian.  In Sec.~\ref{sec: II}, we show the steps to properly use the RWA and the necessary assumptions to find an analytical solution to the dynamic equation.  We show that only the component of the oscillating field perpendicular to the combined static field affects the dynamics within the RWA.  More importantly, our solution provides the full Cartan-decomposed expression for the gate, including the local operations needed to convert the gate to a purely nonlocal controlled phase gate.  We also calculate the local invariants \cite{Makhlin2002}, which completely characterize the nonlocal properties of the evolution operator and the logical gates that it can generate. In Sec.~\ref{sec: III}, we apply our results to a recent experiment with capacitively coupled singlet-triplet qubits \cite{Nichol2016}.
\section{Analytical solution}\label{sec: II}
We consider two qubits with a static Ising coupling, $\alpha$. The local part of the Hamiltonian has an uncontrolled static term along $x$, $h$, and a control field along $z$ oscillating with amplitude $j$ about an average value $J$. We have capacitively coupled singlet-triplet qubits \cite{Shulman2012,Calderon-Vargas2015a} in mind, but our results are applicable to any system with Ising coupling and nonperpendicular local fields, e.g., coupled flux qubits \cite{Plourde2004}. The evolution operator, $U$, is then determined by
\begin{equation}\label{eq:U in lab frame}
i\dot{U}=\left[\sum_{i=1}^2 \left(\frac{J_{i}+j_{i}\cos[\omega_i t]}{2}\sigma_{Z}^{(i)} + \frac{h_i}{2}\sigma_{X}^{(i)}\right) + \alpha \sigma_{ZZ}\right] U,
\end{equation}
where $\sigma_{ij}\equiv \sigma_i^{(1)} \otimes \sigma_j^{(2)}$. It is advantageous to perform a local frame rotation such that the $x$ axis lies along the vector sum of the combined local static parts of the Hamiltonian, so by defining $U_1$ by
\begin{equation}\label{eq:U1def}
U = \exp\left[\frac{i}{2}\sum_{i=1}^2 \phi_i\sigma_{Y}^{(i)}\right] U_1 \exp\left[-\frac{i}{2}\sum_{i=1}^2 \phi_i\sigma_{Y}^{(i)}\right]
\end{equation}
with $\phi_i\equiv \arctan\frac{J_i}{h_i}$, Eq.~\eqref{eq:U in lab frame} becomes
\begin{multline}\label{eq:U_1}
i\dot{U_1}=\bigg[\sum_{i=1}^2 \left(\frac{\Omega_i^2+J_i j_i\cos[\omega_i t]}{2\Omega_i}\sigma_{X}^{(i)}
+2\chi_i\cos[\omega_i t]\sigma_{Z}^{(i)}\right)\\
 +\frac{\alpha}{\Omega_1\Omega_2}\left(J_1 J_2\sigma_{XX}+J_1 h_2\sigma_{XZ}+ J_2 h_1 \sigma_{ZX}+h_1 h_2\sigma_{ZZ}\right)\bigg]U_1,
\end{multline}
where $\Omega_i\equiv \sqrt{J_i^2+h_i^2}$ is the local total energy splitting and $\chi_i\equiv h_i j_i/4\Omega_i$ is the Rabi frequency.  Below we present approximate solutions for various cases.
\subsection{Similar qubits with near-resonant driving}\label{subsec:resonant}
\begin{figure}
  \centering
  \includegraphics[trim=0cm 5cm 0cm 5cm, clip=true,width=8.5cm, angle=0]{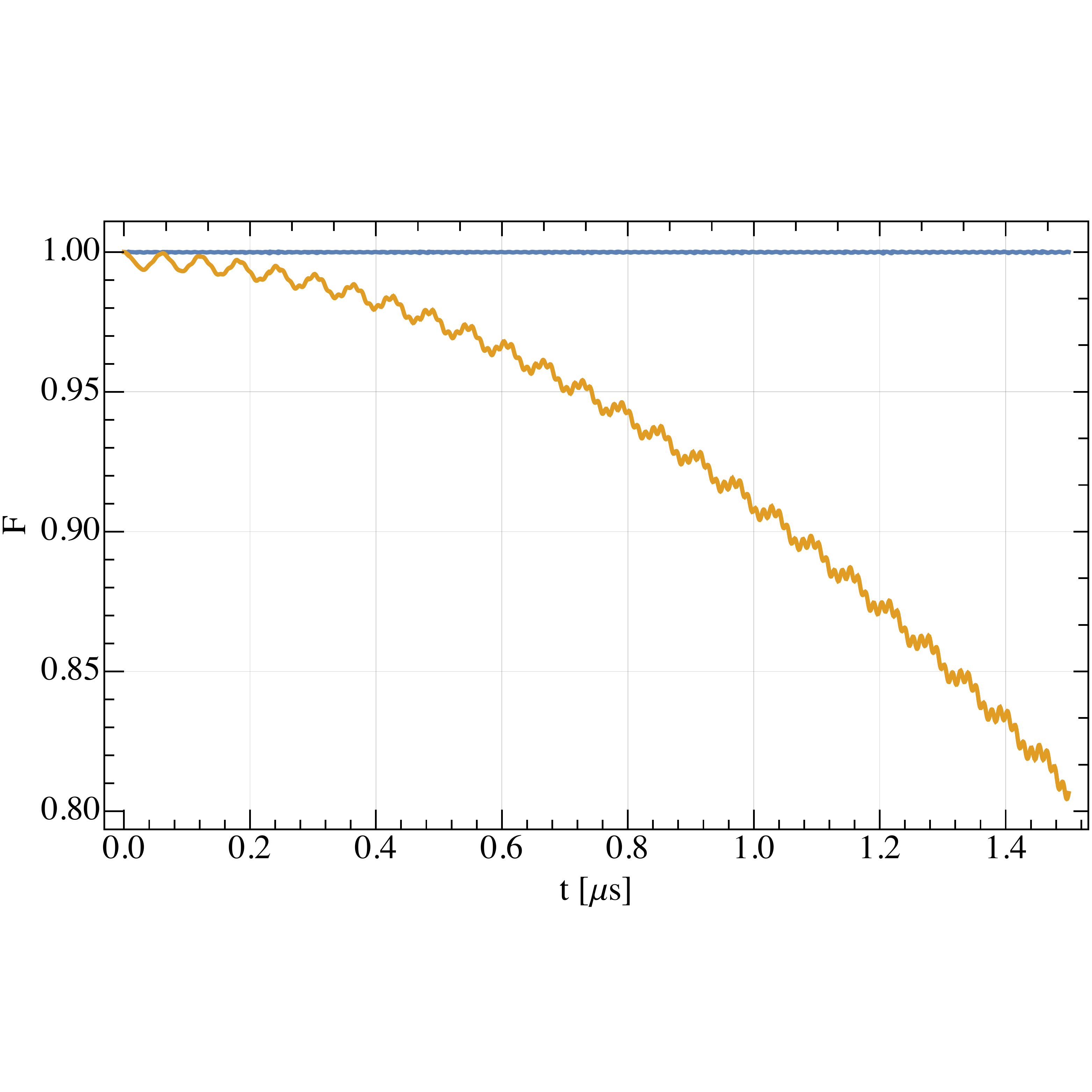}
  \caption{Overlap, $F=\frac{1}{4} |\text{Tr}\left(U^{\dagger}U_{\text{num}}\right)|$, between the approximate evolution operator $U$ and the exact evolution operator $U_{\text{num}}$ obtained numerically.  The straight (oscillatory) curve(s) denote results after one (two) round(s) of RWA \eqref{eq:U_after1RWA_labframe} [\eqref{eq:U_labframe_zz}] using the parameters given in Sec.~\ref{sec: III}.}\label{fig:tracedistance}
\end{figure}

Consider the case of qubits with similar energy splittings and nearly resonant control fields with the same driving frequency, $\omega_1=\omega_2\equiv\omega\sim\Omega_1\sim\Omega_2$.  First we transform to a rotating frame,
\begin{equation}\label{eq:U2def}
U_1=\exp\left[-i \frac{\omega t}{2}\sum_{i=1}^2 \sigma_{X}^{(i)}\right]U_2.
\end{equation}
Typically, one would be interested in producing an entangling gate with duration $T\geq 1/\alpha \gg 1/\omega$, so it is a good approximation to coarse-grain time average over a timescale $\tau \sim 2\pi/\omega$.  In this RWA, terms that go as $e^{\pm i\omega t}$ and $e^{\pm 2i\omega t}$ drop out.  (Note that this cannot be done in the lab frame because the evolution operator itself is rapidly varying there due to the large static term in the lab frame Hamiltonian.  It is safe in the rotating frame, as long as the driving is near resonance and relatively weak, i.e., $\omega \gg \{|\Omega_i-\omega|, |\chi_i|\}$.)  Then the rotating frame Hamiltonian is time-independent and can be directly solved by exponentiation
\begin{multline}\label{eq:U2}
U_2=\exp\left[-it\bigg(\sum_{i=1}^2\left(\frac{\Omega_i-\omega}{2}\sigma_{X}^{(i)}
+\chi_i \sigma_{Z}^{(i)}\right)\right.\\
\left.+\frac{\alpha}{2 \Omega_1 \Omega_2} \left(2 J_1 J_2 \sigma_{XX}
+ h_1 h_2 \left(\sigma_{YY}+\sigma_{ZZ}\right)
\right)\bigg)\right].
\end{multline}
Note that only the perpendicular oscillating terms, $\chi_i$, in Eq.~\eqref{eq:U_1} survive the RWA; the $J_i j_i$ terms do not contribute.\\
\indent Combining with \cref{eq:U1def,eq:U2def} gives the full evolution operator in the lab frame:
\begin{multline}\label{eq:U_after1RWA_labframe}
U=\exp\left[\frac{i}{2}\sum_{i=1}^2 \phi_i\sigma_{Y}^{(i)}\right]\exp\left[-i \frac{\omega t}{2}\sum_{i=1}^2 \sigma_{X}^{(i)}\right]\\
\times\exp\left[-it\bigg(\sum_{i=1}^2\left(\frac{\Omega_i-\omega}{2}\sigma_{X}^{(i)}
+\chi_i \sigma_{Z}^{(i)}\right)\right.\\
\left.+\frac{\alpha}{2 \Omega_1 \Omega_2} \left(2 J_1 J_2 \sigma_{XX}
+ h_1 h_2 \left(\sigma_{YY}+\sigma_{ZZ}\right)
\right)\bigg)\right]\\
\times\exp\left[-\frac{i}{2}\sum_{i=1}^2 \phi_i\sigma_{Y}^{(i)}\right].
\end{multline}
This approximation is quite good for typical parameters, as shown in Fig.~\ref{fig:tracedistance} where the analytical approximation is compared to a numerical solution of the differential equation using {\sc mathematica}'s {\sc ndsolve} for a specific set of values.
However, by sacrificing a bit of precision, we can get a simpler solution that makes the entanglement dynamics more clear.  Define another rotating frame,
\begin{equation}\label{eq:U3def}
  U_2=\exp\left[-it\sum_{i=1}^2 \chi_i\sigma_{Z}^{(i)}\right]U_3.
\end{equation}
In this new frame there are again time-varying terms that go as $e^{\pm i 2\chi_i t}$, $e^{\pm i2(|\chi_1|+|\chi_2|)t}$, and $e^{\pm i2(|\chi_1|-|\chi_2|)t}$. We perform another round of RWA, averaging over an arbitrary timescale bounded by $\pi/|\chi_i|<\tau \ll 1/\alpha$, implicitly assuming that $|\chi_i|\gg\{\alpha,|\Omega_i-\omega|\}$.  (Note that this assumption is more readily satisfied when $h_i > J_i$, as in Ref.~\cite{Nichol2016}, than vice versa.)  The first two types of time-varying terms drop out, but the $e^{\pm i2(|\chi_1|-|\chi_2|)t}$ terms can only be handled in two cases: (a) $||\chi_1|-|\chi_2||\ll 1/T\sim\alpha$, in which case they are approximately constant, or (b) $||\chi_1|-|\chi_2||> 2\pi/\tau$, i.e., the difference in Rabi frequencies is not much less than the smallest one, in which case they drop out along with the other oscillating terms. Thus, for $||\chi_1|-|\chi_2||\ll 1/T\sim\alpha$ we have
\begin{equation}\label{eq:U3equal}
U_3=\exp\left[\frac{-i\alpha t}{2}\left(\frac{h_1 h_2+2J_1 J_2}{2\Omega_1\Omega_2}\left(\sigma_{XX}+\sigma_{YY}\right)+ \frac{h_1 h_2}{\Omega_1\Omega_2}\sigma_{ZZ}\right)\right].
\end{equation}
and the evolution operator in the lab frame can again be obtained by combining \cref{eq:U1def,eq:U2def,eq:U3def,eq:U3equal}.
\indent On the other hand, for $||\chi_1|-|\chi_2|| > 2\pi/\tau$, we have
\begin{equation}\label{eq:U3diff}
U_3=\exp\left[-i t \frac{\alpha h_1 h_2}{2\Omega_1\Omega_2}\sigma_{ZZ}\right]
\end{equation}
and the corresponding evolution operator in the lab frame is
\begin{multline}\label{eq:U_labframe_zz}
U=\exp\left[\frac{i}{2}\sum_{i=1}^2 \phi_i\sigma_{Y}^{(i)}\right] \exp\left[-i \frac{\omega t}{2}\sum_{i=1}^2 \sigma_{X}^{(i)}\right]\\
\times \exp\left[-it\sum_{i=1}^2 \chi_i\sigma_{Z}^{(i)}\right]\exp\left[-i t \frac{\alpha h_1 h_2}{2\Omega_1\Omega_2}\sigma_{ZZ}\right] \\
\times \exp\left[-\frac{i}{2}\sum_{i=1}^2 \phi_i\sigma_{Y}^{(i)}\right].
\end{multline}
The accuracy of this approximation, Eq.~\eqref{eq:U_labframe_zz}, is shown in Fig.~\ref{fig:tracedistance}.\\
\indent The nonlocal properties of the evolution operator, i.e., its entangling properties, are characterized by the operator's local invariants, i.e., quantities that remain invariant under local rotations. In a nutshell, the local invariants are the coefficients of the characteristic polynomial of the symmetric matrix $m(U)$, defined as $m(U)=(Q^{\dagger}U Q)^{\mathrm{T}}Q^{\dagger}U Q $ (where $Q$ denotes the transformation of the matrix $U$ from the logical basis into the Bell basis), and whose spectrum is invariant under local operations \cite{Makhlin2002}. The local invariants equations are
\begin{equation}
\begin{aligned}
G_1=&\frac{\mathrm{tr}^2[m(U)]}{16 \ \mathrm{det}[U]},\\
G_2=&\frac{\mathrm{tr}^2[m(U)]-\mathrm{tr}[m^2(U)]}{4 \ \mathrm{det}[U]}.
\end{aligned}
\end{equation}
\indent Using \cref{eq:U3equal} for $||\chi_1|-|\chi_2||\ll 1/T\sim\alpha$ we obtain the evolution operator's local invariants
\begin{equation}\label{eq: local_inv_equal}
\begin{aligned}
G_1=&\frac{1}{16}\bigg(\cos\left[\frac{4(h_1h_2+J_1J_2) \alpha t}{\Omega_1\Omega_2}\right]+ 6\cos\left[\frac{2h_1 h_2\alpha t}{\Omega_1\Omega_2}\right]\\
&+\cos\left[\frac{4J_1 J_2\alpha t}{\Omega_1 \Omega_2}\right]+8\cos\left[\frac{(h_1 h_2+2J_1 J_2)\alpha t}{\Omega_1 \Omega_2}\right]\bigg)\\
&+\frac{i}{16} \bigg(-2 \sin\left[\frac{2 h_1 h_2\alpha t}{\Omega_1\Omega_2}\right]-\sin \left[\frac{4 J_1J_2 \alpha  t}{\Omega_1\Omega_2}\right]\\
&+\sin \left[\frac{4  (h_1h_2+J_1J_2)\alpha  t}{\Omega_1\Omega_2}\right]\bigg),\\
G_2=&\cos\left[\frac{2h_1 h_2\alpha t}{\Omega_1 \Omega_2}\right]+2\cos\left[\frac{(h_1 h_2+2J_1 J_2)\alpha t}{\Omega_1 \Omega_2}\right].
\end{aligned}
\end{equation}
Alternatively, using Eq.~\eqref{eq:U_labframe_zz} for $||\chi_1|-|\chi_2|| > 2\pi/\tau$, we calculate the following local invariants
\begin{equation}\label{eq: local_inv_diff}
G_1=\cos^2\left[\frac{h_1 h_2\alpha t}{\Omega_1 \Omega_2}\right],\
G_2=2+\cos\left[\frac{2h_1 h_2\alpha t}{\Omega_1 \Omega_2}\right].
\end{equation}
A gate locally equivalent to a {\sc cnot} (or, equivalently, {\sc cphase}) has $G_1=0,~G_2=1$.  Here that is generated when $\frac{h_1 h_2\alpha t}{\Omega_1 \Omega_2}=\frac{\pi}{2}$.
The entangling power, which quantifies the average produced entanglement, is $ep(U)=2/9 \ \left[1-\vert  G_1\vert\right]$ \cite{Balakrishnan2010}, and perfect entangling gates \cite{Makhlin2002,Zhang2003}, i.e., gates that can produce a maximally entangled state from an unentangled one, have $1/6\leq ep(U)\leq 2/9$ and $-1\leq G_2 \leq 1$ \cite{Balakrishnan2010}.
\subsection{Dissimilar qubits with near-resonant driving}\label{subsec: dissimilar}
We can also consider the case when the qubits have very different energy splittings, $\left|\Omega_1 - \Omega_2\right|\gg \{\alpha,\chi_i\}$.  This may occur, for example, in an array of singlet-triplet qubits in silicon where the $h_i$ terms are produced by a fixed, asymmetrical micromagnet instead of being tunable by dynamical nuclear spin polarization.  There one might easily have $|h_1-h_2| \sim$ GHz \cite{Yoneda2015,Noiri2016}. We again transform to a rotating frame,
\begin{equation}\label{eq:U2def2}
U_1=\exp\left[-i \frac{t}{2}\sum_{i=1}^2 \omega_i\sigma_{X}^{(i)}\right]U_2.
\end{equation}
If there exists a timescale $\tau$ such that $\{2\pi/\text{min}\{\omega_i\}, 2\pi/|\omega_1-\omega_2| \}< \tau \ll 1/\alpha$, time-averaging gets rid of terms that go as $e^{\pm i\omega_i t}$, $e^{\pm 2i\omega_i t}$, $e^{\pm i(\omega_1+\omega_2) t}$, and $e^{\pm i(\omega_1-\omega_2) t}$.  (We are again implicitly assuming the driving is near resonance and weak.)  This time we obtain
\begin{equation}\label{eq: Dissimilar First approx}
U_2=\exp\left[-it\left(\sum_{i=1}^2\left(\frac{\Omega_i-\omega_i}{2}\sigma_{X}^{(i)}
+\chi_i\sigma_{Z}^{(i)}\right)
+\frac{\alpha J_1 J_2}{\Omega_1 \Omega_2}\sigma_{XX}\right)\right].
\end{equation}
Using the same arguments as before, we can go to the frame defined by Eq.~\eqref{eq:U3def} and apply another round of RWA to simplify the approximate solution, but now in the case of dissimilar Rabi frequencies, the nonlocal part of the evolution does not survive the averaging and one is left with a local gate.  On the other hand, in the case of similar Rabi frequencies, $||\chi_1|-|\chi_2||\ll 1/T\sim\alpha$, one has
\begin{equation}\label{eq: Dissimilar second approx_equal}
U_3=\exp\Big[-i \alpha t\frac{J_1 J_2}{2\Omega_1\Omega_2}(\sigma_{XX}+\sigma_{YY})\Big].
\end{equation}
Interestingly, comparing Eq.~\eqref{eq: Dissimilar second approx_equal} to Eq.~\eqref{eq:U3diff}, the character of the effective nonlocal coupling has changed from Ising to $XY$, and the coupling strength itself is only different by interchanging $h_i \leftrightarrow J_i$.  Although one might think that makes the nonlocal term depend more sensitively on fluctuations in $J_i$, taking a derivative shows that that is not the case.  We leave a more detailed analysis of the sensitivity of the various parameter regimes to a future work.
The local invariants of this evolution operator are
\begin{equation}\label{eq: local_inv_dissimilar}
G_1=\cos^2\left[\frac{J_1 J_2\alpha t}{\Omega_1 \Omega_2}\right],\
G_2=1+2\cos\left[\frac{2J_1 J_2\alpha t}{\Omega_1 \Omega_2}\right],
\end{equation}
and the operator can generate an i{\sc swap} gate $(G_1=0,~G_2=-1)$ when $\frac{2J_1 J_2\alpha t}{\Omega_1 \Omega_2}=\frac{\pi}{2}$, where i{\sc swap} is equivalent to a combination of {\sc swap} and {\sc cnot} up to single-qubit operations.
\section{Example: coupled singlet-triplet qubits in GaAs}\label{sec: III}
\begin{figure}
  \centering
  \includegraphics[trim=0cm 1cm 0cm 1cm, clip=true,width=8.5cm, angle=0]{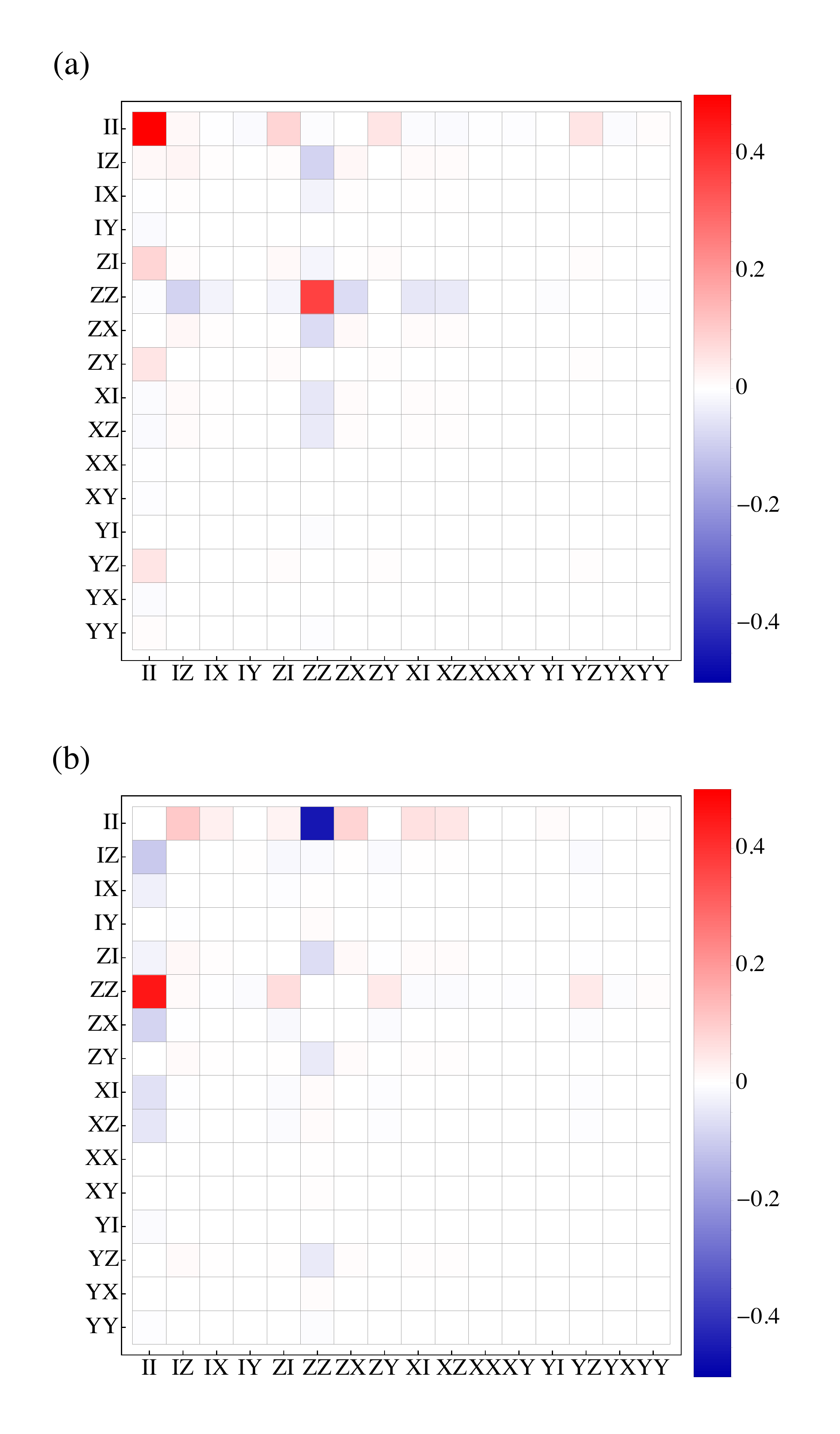}
  \caption{(a) Real and (b) imaginary components of the process matrix at $t\sim 616~\mathrm{ns}$ using the parameters given in Sec.~\ref{sec: III}.}\label{fig:2}
\end{figure}
\begin{figure}
  \centering
  \includegraphics[trim=0cm 7.5cm 0cm 0cm, clip=true,width=8.5cm, angle=0]{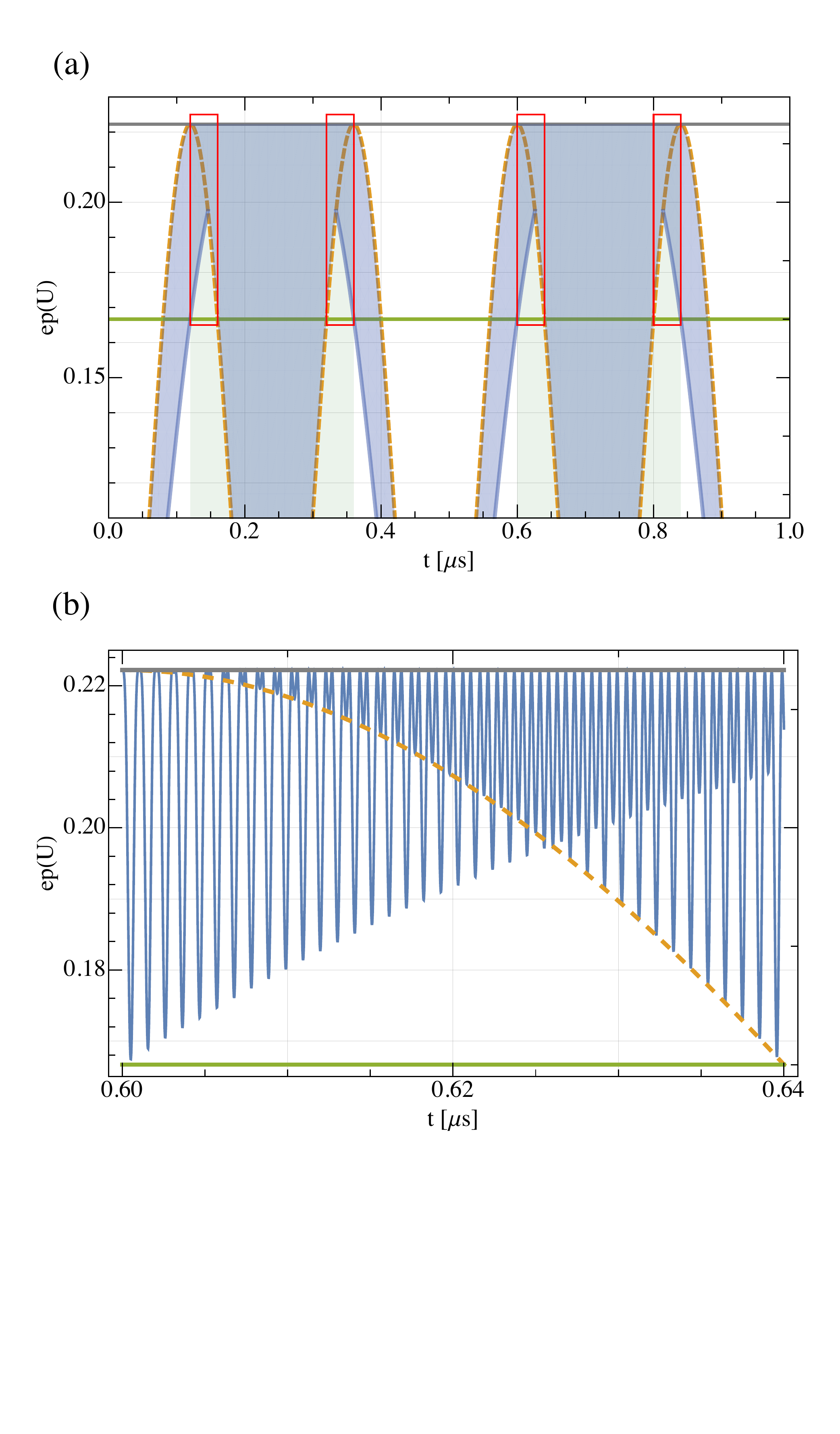}
  \caption{(a) Entangling power vs. time for the evolution operator $\mathcal{U}(t)=U(\chi_i,t/2)U(-\chi_i,t/2)$, using the parameters given in Sec.~\ref{sec: III}. The rapidly oscillating entangling power is marked by the blue shaded region. The light green shaded areas are intervals where $-1\leq G_2 \leq 1$ according to Eq.~\eqref{eq: local_inv_diff_rotary}. Perfect entangling gates are generated in the subsets of the shaded regions where the entangling power is above $1/6$ (horizontal green line). The vertical red rectangles mark the time intervals where the likelihood of generating a maximally entangling gate is maximum. (b) A close-up of (a) around $t\sim 616~\mathrm{ns}$.}\label{fig:3}
\end{figure}
\indent We apply the results obtained in the previous section to two capacitively coupled singlet-triplet qubits.  The experimental setup consists of two similar qubits with near-resonant driving frequency where, using the experimentally reported values of Ref.~\cite{Nichol2016,PrivCommHarvey}, the static components of the exchange interaction energies at each qubit are $J_1/2\pi=266.4\mathrm{MHz}$ and $J_2/2\pi=320\mathrm{MHz}$, the respective magnetic field differences are $h_1/2\pi=922.3$MHz and $h_2/2\pi=905.1$MHz, the driving frequency in each qubit is $\omega_1=\omega_2\equiv \omega= (2\pi) 960\mathrm{MHz}\sim \Omega_1\sim\Omega_2$, oscillation amplitudes are in the range $25\mathrm{MHz}\leq j_i/2\pi \leq 70\mathrm{MHz}$ and approximately different by a factor of two \cite{PrivCommNichol}, and the coupling strength is in the range $0.4\mathrm{MHz}\leq \alpha/2\pi \leq 2.5\mathrm{MHz}$, corresponding to entangling times of hundreds of ns. Note that the assumptions made in the first round of RWA leading to Eq.~\eqref{eq:U_after1RWA_labframe}, $\omega\gg\chi_i,\alpha$, are clearly valid for these values, and that the factor of two difference in experimental Rabi frequencies means that the assumption made in the second round of RWA leading to Eq.~\eqref{eq:U_labframe_zz} is also valid.\\
\indent The use of rotary echo \cite{Solomon1959a}, as in Ref.~\cite{Nichol2016}, is also included in our analysis. The rotary echo is applied simultaneously to both qubits, giving an evolution operator $\mathcal{U}(t)=U(\chi_i,t/2)U(-\chi_i,t/2)$.\\
\indent According to the concurrence plot in Ref.~\cite{Nichol2016}, the gate time of the first peak is around $600\mathrm{ns}$, so for the sake of comparison we focus on gates that can be generated close to that time, although a maximally entangling gate could also be generated earlier.\\
\indent Using Eq.~\eqref{eq:U_labframe_zz} and rotary echo, we perform quantum process tomography \cite{Chuang1997} to calculate the process matrix (which characterizes the action of a process on the components of the density matrix) at different times, coupling strengths, and Rabi frequencies in order to compare to the measured process matrix reported in Ref.~\cite{Nichol2016}. In Fig.~\ref{fig:2}, we show the process matrix at a time equal to $615.7\mathrm{ns}$, a coupling strength $\alpha=2\pi\times 2.3\mathrm{MHz}$, and $j_1 =2\pi\times 69.3\mathrm{MHz},~j_2=2\pi\times36\mathrm{MHz}$ \cite{Note}. These values produce a positive process matrix $\chi$ relatively close to the not completely positive process matrix $\chi_{exp}$ reported in Ref.~\cite{Nichol2016} [$\mathrm{Tr}(\chi~\chi_{exp})=86\%$]. Note that in Ref.~\cite{Nichol2016}, after using a maximum likelihood estimation process to ensure a completely positive process matrix, a brute-force numerical search over all possible completely positive process matrices generated by an Ising coupling matches the experimental result with at best $87\%$ fidelity \cite{Nichol2016}. Thus, in the absence of the experimental noise causing deviations from positivity, our theory captures the dynamics well. Moreover, the process matrix presented here is uniquely generated by a given set of experimental parameters and does not require a search over single-qubit rotations as previous modeling did \cite{Nichol2016}.\\
\indent As a consequence of the rotary echo, the local invariants and entangling power of the evolution operator now depend on the driving frequency $\omega$, in contrast to Eq.~\eqref{eq: local_inv_diff}, and are given by
\begin{equation}\label{eq: local_inv_diff_rotary}
\begin{aligned}
G_1=&\frac{1}{16}\left(1-\cos[\omega t]+(3+\cos[\omega t])\cos\left[\frac{h_1 h_2\alpha t}{\Omega_1 \Omega_2}\right]\right)^2,\\
G_2=&\frac{1}{2}\Bigg(3+\cos\left[\frac{2h_1 h_2\alpha t}{\Omega_1 \Omega_2}\right]\\
&+\cos\left[\frac{h_1 h_2\alpha t}{\Omega_1 \Omega_2}\right]\left(2-4\cos[\omega t]\sin^2\left[ \frac{h_1h_2\alpha t}{2\Omega_1\Omega_2}\right] \right)\Bigg).
\end{aligned}
\end{equation}
We can identify the time frames where maximally entangling gates are generated by plotting the entangling power versus time, as shown in Fig.~\ref{fig:3}. During the first microsecond there are two shaded time intervals in Fig. ~\ref{fig:3} where perfect entangling gates are produced, but the entangling power oscillates rapidly. Nonetheless, there are regions within the shaded time intervals where the average likelihood of generating a maximally entangling gate is larger. These regions are marked by red vertical rectangles in Fig.~\hyperref[fig:3]{\ref*{fig:3}(a)} and are delimited by the points where the slowly oscillating envelope (dashed curve) of the entangling power, calculated with Eq.~\eqref{eq: local_inv_diff}, is equal to 2/9 or 1/6. In these regions the evolution operator can, in principle, produce a maximally entangling gate despite possible noise effects. The gate time associated with the process matrix of Fig.~\ref{fig:2}, $t=615.7\mathrm{ns}$, falls within one of these intervals [Fig.~\hyperref[fig:3]{\ref*{fig:3}(b)}]. In fact, the evolution operator, whose local invariants at that time are $\{G_1\sim0.03,~G_2\sim1.06\}$, is equivalent to a {\sc cphase} gate up to single-qubit rotations with a fidelity of $\sim 99.2\%$. This fidelity is in the absence of noise, which will of course reduce the fidelity.  Furthermore, considering the initial Hamiltonian, Eq.~\eqref{eq:U in lab frame}, numerically with the parameters given in this section and using rotary echo we find that the corresponding evolution operator is indeed equivalent to a {\sc cphase} gate up to single-qubit rotations with a fidelity of $\sim 99.1\%$. This shows that our approximate analytical solution conveys with high precision the nonlocal properties of the system.\\
\indent Our results are in broad agreement with Ref.~\cite{Nichol2016}, where the experimental process matrix was equivalent to a {\sc cphase} gate up to single-qubit rotations with $90\%$ fidelity (or $87\%$ if one enforces complete positivity).  One may understand why the experimentally reported {\sc cphase} fidelity was only $\sim 90\%$ instead of the $\sim 99\%$ we calculate.  The main difference, of course, is that we have not included the effects of noise in our calculations, a detailed noise analysis lying well beyond the scope of this paper. However, by considering that the $T_2^{\text{echo}}$ decoherence time was longer than the experimental entanglement time by about a factor of six for the Rabi frequencies in use we can roughly estimate the magnitude of the effect of the noise.  Note that the reported $T_2^{\text{echo}}$ is a single-qubit decoherence time, but it should be similar for two qubits because charge noise gives rise to fluctuations in $J$ and $j$, in turn producing fluctuations in the Rabi frequency $\chi$ which causes random local phases to accumulate in Eq.~\eqref{eq:U_labframe_zz}, while the fluctuations in coupling strength are negligible in comparison \cite{Shulman2012}.  Thus, the entangling part of the gate is relatively unaffected, since the perturbation in the nonlocal phase is proportional to $J \delta J/\Omega^2$ and $\Omega>>J$.  Only the accompanying local rotations are strongly affected.  So then, considering charge noise with a power spectral density of $1/f^{0.7}$, one might expect fidelity when using the rotary echo to be roughly comparable to $e^{-2\left(t/T_2^{\text{echo}}\right)^{1.7}} \sim 90\%$, where the factor of two in the exponent is because independent local dephasing is occurring on both qubits.  The experimentally reported {\sc cphase} fidelity of $\sim 90\%$ is not surprising then, even apart from other possible effects such as imperfections in the rotary echoes, control calibration, and tomography.
\section{Conclusions}
We have presented an approximate analytical solution to the dynamic equation of an Ising-coupled two-qubit system whose oscillating and static fields are not perpendicular. This solution has been obtained by an appropriate chain of local transformations and proper application of the RWA. Moreover, by isolating the nonlocal content of the evolution for four cases, depending on whether the qubits' energy splittings and Rabi frequencies are similar or not, we calculated the local invariants, which determine the type of logical gates the system can generate.\\
\indent We applied our results to a recent experimental work on capacitively coupled singlet-triplet qubits in GaAs. Our solution gives new insight into the type of entangling gates which are being generated, predicts the presence of other perfect entangling gates at shorter times, and presents regions where the two-qubit evolution operator can produce at least minimally entangling gates regardless of possible perturbations. Furthermore, our discussion of a case with highly asymmetric qubit energy splittings is relevant for future experiments in silicon where dynamical nuclear spin polarization is not available to tune the $h_i$ fields. Future work will explore the sensitivity of the entangling gates to various physical noise sources. However, the results presented in this work are immediately relevant to ongoing experiments with capacitively coupled singlet-triplet qubits as well as any system with a similar form of Hamiltonian.
\section{Acknowledgments}
We thank John Nichol and Shannon Harvey for valuable discussions. J.P.K. acknowledges support by the Army Research Office (ARO) under Grant No. W911NF-17-1-0287, and F.A.C. acknowledges support by the National Science Foundation under Grant No. 1620740.

\bibliographystyle{apsrev4-1}

\bibliography{library}

\begin{thebibliography}{25}
\bibitem [{\citenamefont {{For these parameters the fidelity after two rounds
  of RWA is 95\%}}()}]{Note}%
  \BibitemOpen
  \bibfield  {author} {\bibinfo {author} {\bibnamefont {{For these parameters
  the fidelity after two rounds of RWA is 95\%}}},\ }\href@noop {} {\bibinfo
  {journal} {{which improves for larger differences in driving amplitudes}}\
  }\BibitemShut {NoStop}%
\end{thebibliography}

\end{document}